\documentclass[times,3p,a4paper,number]{elsarticle}

\usepackage{amsmath} 
\usepackage{amsfonts}
\usepackage{graphicx} 
\usepackage{color}
\usepackage{subfigure}
\usepackage[latin2]{inputenc}
\usepackage{textcomp} 
\usepackage{multirow}
\usepackage{rotating}

\usepackage{floatpag}
\rotfloatpagestyle{empty}

\usepackage[normalem]{ulem}
\renewcommand{\uline}{}

\journal{Physica A}

\begin{document}

\begin{frontmatter}

\title{Compensating asynchrony effects in the calculation of financial correlations}

\author{Michael C. M\"unnix\corref{cor1}}
\ead{michael@muennix.com}
\author{Rudi Sch\"afer\corref{}}
\author{Thomas Guhr\corref{}}
\cortext[cor1]{Corresponding author. Tel.: +49 203 379 4727; Fax: +49 203 379 4732.}

\address{Fakult\"at f\"ur Physik, Universit\"at Duisburg-Essen, Germany}

\begin{abstract}
We present a method to compensate statistical errors  in the calculation of correlations on asynchronous time series. The method is based on the assumption of an underlying time series.
We set up a model and apply it to financial data to examine the decrease of calculated correlations towards smaller return intervals (Epps effect). We show that this statistical effect is a major cause of the Epps effect. Hence, we are able to quantify and to compensate it using only trading prices and trading times.
\end{abstract}


\end{frontmatter}

\section{Introduction}
The decrease of calculated correlations in financial data towards smaller return (or ``sampling''-) intervals has been of interest since Epps discovered this phenomenon in 1979 \cite{epps79}. Ever since, this behavior was found in data of different stock exchanges  \cite{Bonanno01, kwapien04, tumminello07,zebedee09} and foreign exchange markets \cite{lundin98, muthuswamy01}.

Many economists as well as physicists addressed this phenomenon, since a precise calculation of correlations is of major importance for the estimation of financial risk \cite{schaefer07,schaefer08,onnela03}. While the physicists' approach is often to construct a model which offers an explanation for this phenomenon, the standard economy approach is to work on estimators with the aim to suppress the Epps effect. Recently, Hayashi and Yoshida  introduced a cumulative estimator \cite{hayashi05}, only involving returns whose time intervals are overlapping. This estimator has been supplemented with different adjustments, such as bias compensation and \emph{lead-lag} treatment \cite{zebedee09,voev07,griffin06}. A very similar approach on a completly different topic is the ``discrete correlation function'' in astrophysics which was introduced in 1988 by Edelson and Krolik \cite{edelson88}. Other approaches to estimate correlation coefficients involve \emph{Previous-Tick-Estimators} \cite{corsi07,zhang08} or realized kernel functions \cite{barndorff08}.

An extensive study of microscopic causes leading to the Epps effect has been performed by Ren\`o \cite{reno03}, while another work by Tóth et. al. introduce a model for the Epps effect which is based on the phenomenon of lagged correlations \cite{toth09}.

However, certainly miscellaneous mechanisms are contributing to the Epps effect. Thus our approach is different.
First, we will introduce a simple model which offers an explanation for the statistical part of the Epps effect, based on a central assumption of an underlying time series. Secondly, based on that model, we will present an estimator, with which these effects can be compensated. Finally, we will quantify the impact of this phenomenon on the Epps effect in recent empirical data and show that it can be a major cause for the Epps effect, especially when looking at less frequently traded securities.

This paper is organized as followed: In section \ref{statef}, we develop the model for correlations in asynchronous time series. Within the model, we observe a decay of correlations towards smaller return intervals, similar to the Epps effect. We then derive the method to compensate this phenomenon. In section \ref{application}, this method is applied to recent empirical data to estimate the impact of the observed effect on the Epps effect. We discuss the results in section \ref{conclusion}.

\section{Statistical effects in asynchronous time series}
\label{statef}
In section \ref{modeling}, we set up our model and develop a compensation formalism for asynchrony effects in section \ref{compensation}.

\subsection{Model}
\label{modeling}
The central assumption of our model is the existence of an underlying non-lagged time series of prices. The assumption of a finer \cite{toth09} or even continuous \cite{hayashi05,barndorff02,reno03} underlying timescale is a common approach in the estimation of correlations. This approach is also intuitive, as most stocks are traded at several stock exchanges simultaneously. 

\begin{figure}[tb]
\begin{center}
\includegraphics[width=0.60\textwidth]{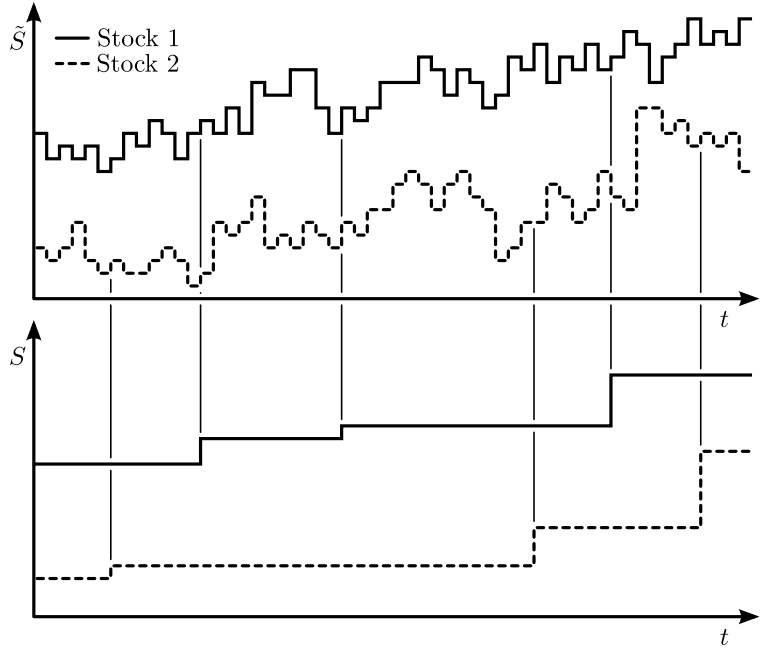}
\caption{Illustration of the model for asynchronous trading times of two stocks. Shown above are the prices $\tilde{S}$ on the underlying timescale. The ``sampling'' of theses prices $\tilde{S}$ to prices $S$ on simulated trading times are shown below.}
\label{fig:model}
\end{center}
\end{figure}

To simulate asynchrony effects we generate an underlying correlated time series using the \emph{Capital Asset Pricing Model} (CAPM) \cite{sharpe64}, which is also known as Noh's model \cite{noh00} in physics,
\begin{equation}
\tilde{r}^{(i)}(t) = \sqrt{c}\,\eta(t) + \sqrt{1-c}\,\varepsilon^{(i)}(t)\ ,
\end{equation}
where $\tilde{r}^{(i)}$ stands for the relative price change, the so-called \emph{return} of the $i$-th stock and $c$ is the correlation coefficient. The random variables $\eta$ and $\varepsilon^{(i)}$ are taken from a compound distribution as observed on market data by Gopikrishnan et. al. with power-law tails and a central Levy distribution (for details see Ref. \cite{gopikrishnan99}).
\uline{We have chosen this approach to keep our model initially as simple as possible. We note, however, that return time series can also be autocorrelated. While first order autocorrelations are  in this context insignificantly small \cite{voit03}, second order autocorrelations or ``volatility clustering'' represent a strong characteristic of return time series and led to the development of autoregressive models, such as GARCH} \cite{engle82,engle01}. \uline{For this reason we also test our compensation in a more realistic setup against a GARCH(1,1) generated time series of underlying returns, given by}

\begin{equation}
\tilde{r}^{(i)}(t) = \sigma^{(i)}(t)\left(\sqrt{c}\,\eta(t) + \sqrt{1-c}\,\varepsilon^{(i)}(t)\right) 
\end{equation}
\uline{with}
\begin{equation}
\left(\sigma^{(i)}(t)\right)^{2} = \alpha_{0}+\alpha_{1} \left(\tilde{r}^{(i)}(t-1)\right)^{2} + \beta_{1}\left(\sigma^{(i)}(t-1)\right)^{2}\ . 
\end{equation}
\uline{The initial parameters of the GARCH process have been chosen as $\alpha_{0}=2.4\cdot10^{-4}$, $\alpha_{1}=0.15$ and $\beta_{1}=0.84$.}

Two return time series $\tilde{r}^{(1)}$ and $\tilde{r}^{(2)}$ are generated representing two correlated stocks. The lengths of these  underlying time series are chosen as $7.2\cdot10^{6}$, $1.44\cdot10^{6}$ and $7.2\cdot10^{5}$ corresponding to a return interval $\Delta\tilde{t}$ on the underlying timescale of 1, 5 and 10 seconds during 1 trading year.

Using these returns and an arbitrary starting price, the underlying price series $\tilde{S}^{(1)}$ and $\tilde{S}^{(2)}$ are calculated \uline{implying a geometric Brownian motion with zero drift and a standard deviation of $10^{-3}$ per time step}. To model the asynchronous trade processes, these prices are sampled independently using exponentially distributed waiting times with average values typical for the stock market (see Fig. \ref{fig:model}). In the following example, we choose the average waiting times as  $\mu^{(1)}=15$ and $\mu^{(2)}=25$ (equivalent to seconds in this example), while the underlying time series were correlated with $c=0.4$.
\begin{figure}[tb]
\begin{center}
\includegraphics[width=0.50\textwidth]{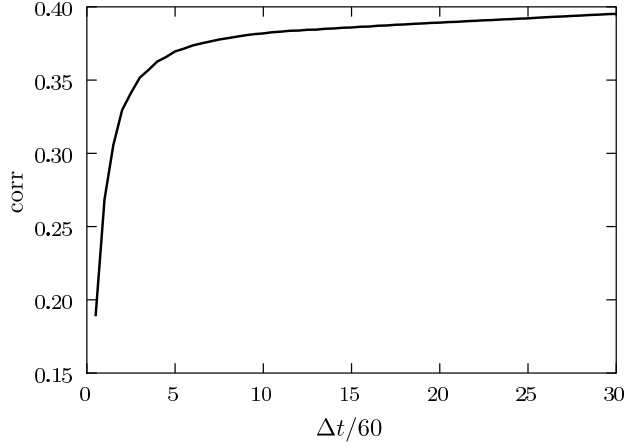}
\caption{Scaling behavior of the correlation coefficient on a simulated asynchronous time series. The length of the underlying time series was set to $7.2\,10^{6}$ ($c=0.4$, $\mu^{(1)}=15$ and $\mu^{(2)}=25$). The correlation coefficient was calculated on return-intervals from  60 data points (corresponding to 1 minute) to \uline{1800 data points (corresponding to 30 minutes).}}
\label{fig:modelcor}
\end{center}
\end{figure}
On the resulting ``macroscopic'' time series, the return between two points in time (of the $i$-th stock) can be calculated as
\begin{equation}
r^{(i)}_{\Delta t}(t)=\frac{S^{(i)}(t+\Delta t) - S^{(i)}(t)}{S^{(i)}(t)}\ ,
\label{return}
\end{equation}
where $S^{(i)}(t)$ denotes the price at time $t$ and $\Delta t$ is the return interval. Between these return time series, we now calculate the correlation coefficient,

\begin{equation}
\mathrm{corr}(r^{(i)}_{\Delta t},r^{(j)}_{\Delta t})=\frac{\left\langle r^{(i)}_{\Delta t}r^{(j)}_{\Delta t} \right\rangle - \left\langle r^{(i)}_{\Delta t} \right\rangle\left\langle r^{(j)}_{\Delta t} \right\rangle}{\sigma^{(i)}_{\Delta t}\sigma^{(j)}_{\Delta t}}\ ,
\label{eq:corrfef}
\end{equation}
where $\langle \ldots \rangle$  denotes the mean value of a time series with length $T$ and where $\sigma$ refers to the standard deviation of the same time series. We note that we refer to the whole time series of returns $r^{(i)}_{\Delta t}$ when the argument $(t)$ is omitted.

When calculating the correlation of returns of the sampled time series $\tilde{S}$ using different return intervals $\Delta t$, the correlation coefficient scales down as shown in Fig \ref{fig:modelcor}. This behavior is very similar to the Epps effect in empirical data. It occurs only because of the asynchrony of the trading times. As this behavior is already observed in this simple setting, we are able to derive a method to compensate it, as the following demonstrates.

\subsection{Compensation}
\label{compensation}
The basic idea of this approach is the following: Due to the asynchrony, each term of the correlation coefficient can be divided into a part which contributes to the correlation and a part which is uncorrelated and therefore lowers the correlation coefficient.

According to the model assumption, the price change during $\Delta t$ is based on price changes on an underlying ``microscopic'' timescale. Thus, the return can also be expressed as a sum of the underlying returns,

\begin{equation}
r^{(i)}(t) = \sum\limits_{j=0}^{N^{(i)}_{\Delta t}(t)}{\tilde{r}^{(i)}(\gamma^{(i)}(t)+j\Delta \tilde{t}})\ .
\label{underl}
\end{equation}
Here $\tilde{r}^{(i)}(t_{i})$ is the return related to $S(t)$ on the underlying time scale of non-overlapping intervals $\Delta \tilde{t}$ (e.g. $1\,\mathrm{second}$) given by

\begin{equation}
\tilde{r}^{(i)}(t+j\Delta \tilde{t}) = \frac{\tilde{S}(t+(j+1)\Delta \tilde{t}) - \tilde{S}(t+j\Delta \tilde{t})}{S(t)}\ .
\end{equation}
The quantity $\gamma^{(i)}(t)$ in equation (\ref{underl}) represents the time of the last trade of the $i$-th stock at time $t$,
\begin{equation}
\gamma^{(i)}(t)=\max(t^{(i)}_{\mathrm{trade}})\Big|_{t^{(i)}_{\mathrm{trade}}\le t}\ .
\end{equation}
When calculating the return for the interval $[t,t+\Delta t]$ of two stocks, the actual price at $t$ and $t+\Delta t$ is generally in the past, more precisely at $\gamma^{(1)}(t)$, $\gamma^{(2)}(t)$ and $\gamma^{(1)}(t+\Delta t)$, $\gamma^{(2)}(t+\Delta t)$. These trading times are distinct for each stock, therefore only a fraction of the underlying prices processed by the return is correlated. 
The number of terms $N^{(i)}_{\Delta t}$ of the sum in equation (\ref{underl}) is given by

\begin{equation}
N^{(i)}_{\Delta t}(t)=\frac{(\gamma^{(i)}(t+\Delta t)-\gamma^{(i)}(t))}{\Delta\tilde{t}}\ .
\label{under}
\end{equation}
We normalize the returns to zero mean and unit variance and indicate them as $g$ and $\tilde{g}$:

\begin{equation}
g^{(i)}_{\Delta t}(t)=\frac{r^{(i)}_{\Delta t}(t)-\langle r^{(i)}_{\Delta t}  \rangle}{\sigma^{(i)}_{\Delta t}}
,\quad
\tilde{g}^{(i)}(t)=\frac{\tilde{r}^{(i)}(t)-\langle \tilde{r}^{(i)}  \rangle}{\tilde{\sigma}^{(i)}}\ .
\end{equation}
In this context, the relation of the returns on both time scales in equation (\ref{underl}) changes to

\begin{equation}
g_{\Delta t}^{(i)}(t) = \sqrt{\frac{\Delta \tilde{t}}{\Delta t}} \sum\limits_{j=0}^{N^{(i)}_{\Delta t}(t)}{\tilde{g}^{(i)}(\gamma^{(i)}(t)+j\Delta \tilde{t}})
- \frac{\langle \tilde{r}^{(i)} \rangle \left( \frac{\Delta t}{\Delta \tilde{t}} - N^{(i)}_{\Delta t}(t)\right)}{\sigma^{(i)}_{\Delta t}}\ ,
\label{eq:appneeded}
\end{equation}
as worked out in appendix \ref{app1}. When using normalized returns, the correlation coefficient of two return time series $r_{\Delta t}^{(1)}$ and $r_{\Delta t}^{(2)}$ (see equation (\ref{eq:corrfef})) simplifies to
\begin{equation}
\mathrm{corr}(r_{\Delta t}^{(1)},r_{\Delta t}^{(2)})=\mathrm{corr}(g_{\Delta t}^{(1)},g_{\Delta t}^{(2)})=
\frac{1}{T}\sum\limits_{j=0}^{T}g_{\Delta t}^{(1)}(t_{j})g_{\Delta t}^{(2)}(t_{j})\ .
\label{corr}
\end{equation}
As the mean value over $T$ of the second term from equation (\ref{eq:appneeded}) is equal to zero, we obtain in terms of the underlying time series
\begin{eqnarray}
\mathrm{corr}(r_{\Delta t}^{(1)},r_{\Delta t}^{(2)})=\frac{1}{T}\sum\limits_{j=0}^{T}\left(
\sum\limits_{k=0}^{N^{(1)}_{\Delta t}(t_{j})}{\tilde{g}^{(1)}(\gamma^{(1)}(t_{j})+k\Delta \tilde{t}})
\sum\limits_{l=0}^{N^{(2)}_{\Delta t}(t_{j})}{\tilde{g}^{(2)}(\gamma^{(2)}(t_{j})+l\Delta \tilde{t}})
\right)
\frac{\Delta \tilde{t}}{\Delta t}\ .
\label{eq:corrunderl}
\end{eqnarray}
\begin{figure}[tb]
\begin{center}
\includegraphics[width=0.60\textwidth]{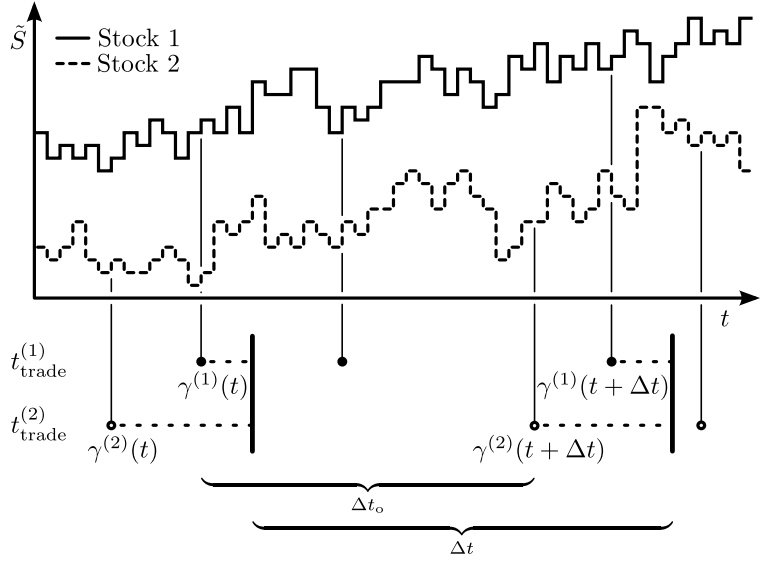}
\caption{Illustration of the overlap $\Delta t_{\mathrm{o}}$.}
\label{fig:overlap}
\end{center}
\end{figure}
\begin{figure}[tb]
\begin{center}
\subfigure[$\Delta t = 150$]{
  \includegraphics[width=0.4\textwidth]{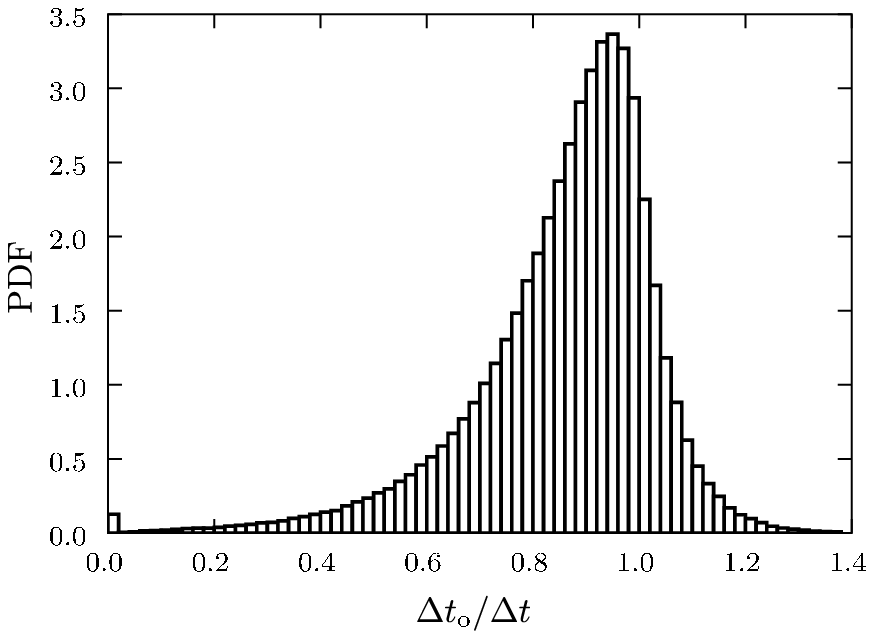}
  \label{fig:overlapsa}
  }
 \subfigure[$\Delta t = 450$]{
   \includegraphics[width=0.4\textwidth]{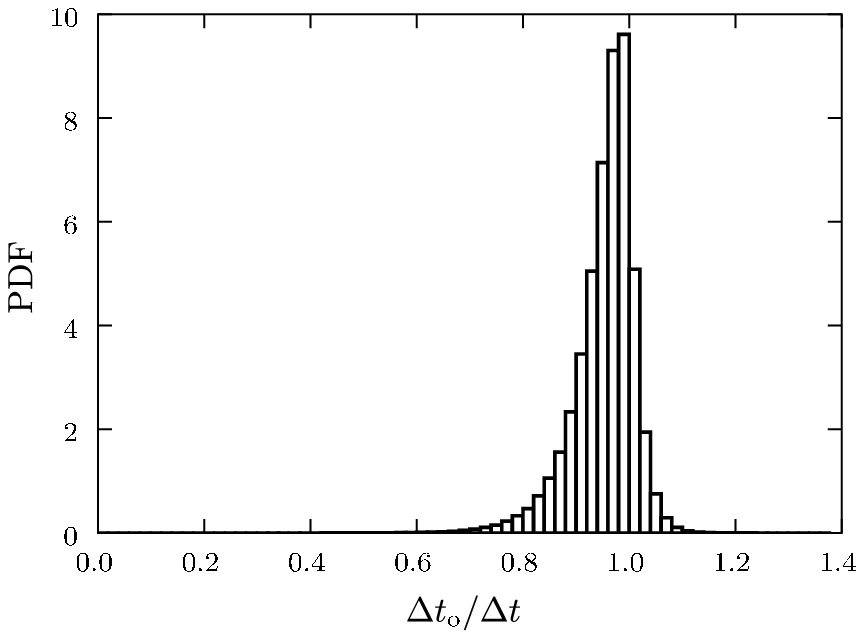}
  }\\
 \subfigure[$\Delta t = 1500$]{
   \includegraphics[width=0.4\textwidth]{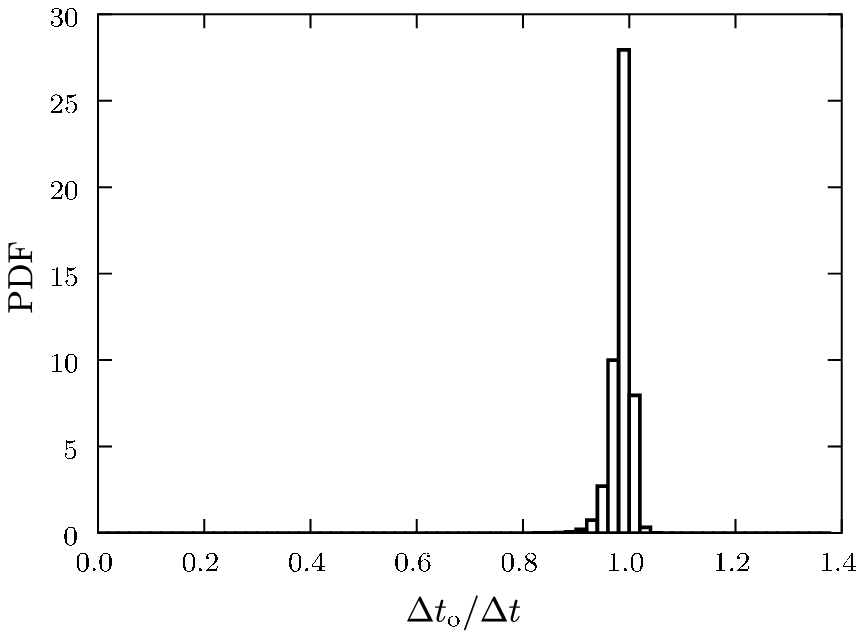}
  }
 \subfigure[Mean overlap]{
   \includegraphics[width=0.4\textwidth]{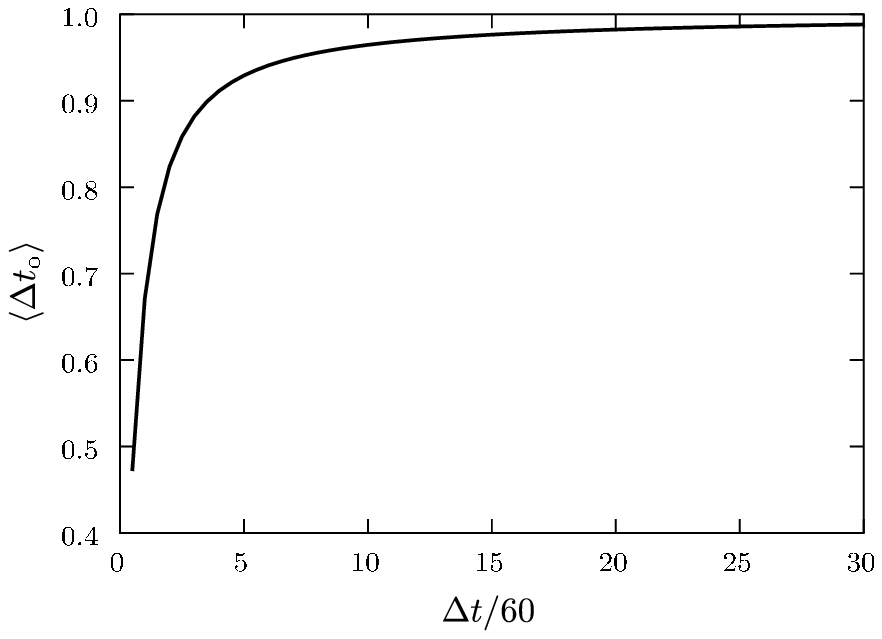}
  }
\caption{Distribution of the overlaps for simulated return intervals $\Delta t=150$ (a), $450$ (b) and $1500$ (c) data points (corresponding to 2.5, 7.5 and 25 minutes). Towards larger return-intervals, the distribution sharpens, as well as its mean converges to 1 (d).}
\label{fig:overlaps}
\end{center}
\end{figure}
As illustrated in Fig. \ref{fig:overlap}, only a subset of the underlying prices $\tilde{S}$ of two prices $S$ share an overlapping time-interval.
Because of this ``overlap'' only  a certain amount $\bar{N}_{\Delta t}(t)$ of the underlying returns is correlated, namely
\begin{equation}
\bar{N}_{\Delta t}(t)=\frac{\Delta t_{\mathrm{o}}(t)}{\Delta \tilde{t}}
\end{equation}
with $\Delta t_{\mathrm{o}}(t)$ being the time interval of the actual overlap,
\begin{equation}
\Delta t_{\mathrm{o}}(t) = \min(\gamma^{(1)}(t+\Delta t),\gamma^{(2)}(t+ \Delta t))-\max(\gamma^{(1)}(t),\gamma^{(2)}(t))\ .
\end{equation}
Each sum can be split up into $N^{(i)}_{\Delta t}-\bar{N}$ terms that are uncorrelated and $\bar{N}$ that are correlated. Thus, equation (\ref{eq:corrunderl}) can be written as:
\begin{eqnarray}
\mathrm{corr}(r_{\Delta t}^{(1)},r_{\Delta t}^{(2)})=
\frac{1}{T}\sum\limits_{j=0}^{T}
\left(\left( \underbrace{\sum\limits_{k=\bar{N}_{\Delta t}(t_{j})+1}^{N^{(1)}_{\Delta t}(t_{j})-\bar{N}_{\Delta t}(t_{j})}{\tilde{g}^{(1)}(t_{k})}}_{\mathrm{async.}} +  \underbrace{\sum\limits_{\bar{k}=0}^{\bar{N}_{\Delta t}(t_{j})}{\tilde{g}^{(1)}(t_{\bar{k}})}}_{\mathrm{sync.}} \right)\right. \nonumber\\
\left.\times\left( \underbrace{\sum\limits_{l=\bar{N}_{\Delta t}(t_{j})+1}^{N^{(2)}_{\Delta t}(t_{j})-\bar{N}_{\Delta t}(t_{j})}{\tilde{g}^{(2)}(t_{l})}}_{\mathrm{async.}} +  \underbrace{\sum\limits_{\bar{l}=0}^{\bar{N}_{\Delta t}(t_{j})}{\tilde{g}^{(2)}(t_{\bar{l}})}}_{\mathrm{sync.}}\right) \frac{\Delta \tilde{t}}{\Delta t}\right) \ ,
\label{decomp}
\end{eqnarray}
where only the sums of synchronous returns are correlated among each other. In this notation, the underlying time series is indexed as $[\tilde{r}^{(i)}(t_{0}),\tilde{r}^{(i)}(t_{1}),\dots\tilde{r}^{(i)}(t_{N^{(i)}_{\Delta t}}))]$, where the returns from $t_{0}$ to $t_{\bar{N}_{\Delta t}}$ are corresponding to the overlap.  When expanding the product, the non-correlated returns converge to zero due to the outer average
\begin{eqnarray}
\mathrm{corr}(r_{\Delta t}^{(1)},r_{\Delta t}^{(2)})&=&\frac{1}{T}\sum\limits_{j=0}^{T}\left(\left( \underbrace{\sum\limits_{k=0}^{\bar{N}_{\Delta t}(t_{j})}{\tilde{g}^{(1)}(t_{k})\tilde{g}^{(2)}(t_{k})}}_{\bar{N}_{\Delta t}(t)\mathrm{corr}_{t_{j}}(\vec{\tilde{r}}_{1},\vec{\tilde{r}}_{2})} +  \underbrace{\dots}_{0}\right) \frac{\Delta \tilde{t}}{\Delta t}\right)\nonumber\\
&=& \frac{1}{T}\sum\limits_{j=0}^{T}\mathrm{corr}_{t_{j}}(\tilde{g}^{(1)},\tilde{g}^{(2)}) \frac{\bar{N}_{\Delta t}(t) \Delta \tilde{t}}{\Delta t} \nonumber\\
&=& \frac{1}{T}\sum\limits_{j=0}^{T}\mathrm{corr}_{t_{j}}(\tilde{g}^{(1)},\tilde{g}^{(2)}) \frac{\Delta{t}_{\mathrm{o}}(t_{j})}{\Delta{t}}\ ,
\label{result}
\end{eqnarray}
where $\mathrm{corr}_{t}$ represents the correlation of the underlying returns corresponding to the interval $[t,t+\Delta t]$.

\begin{figure}[tb]
\begin{center}
\subfigure[Noh]{
\includegraphics[width=0.45\textwidth]{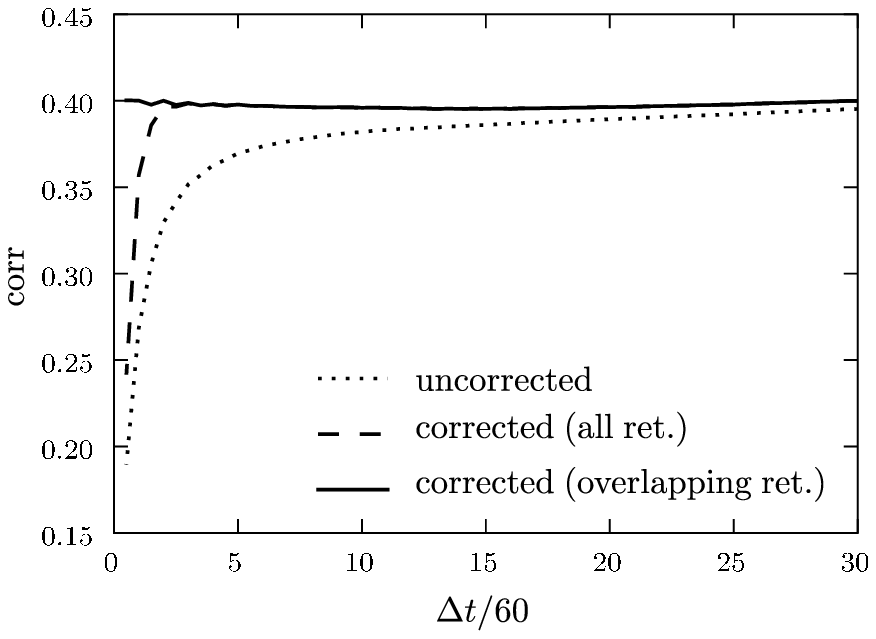}
}
\subfigure[GARCH(1,1)]{
\includegraphics[width=0.45\textwidth]{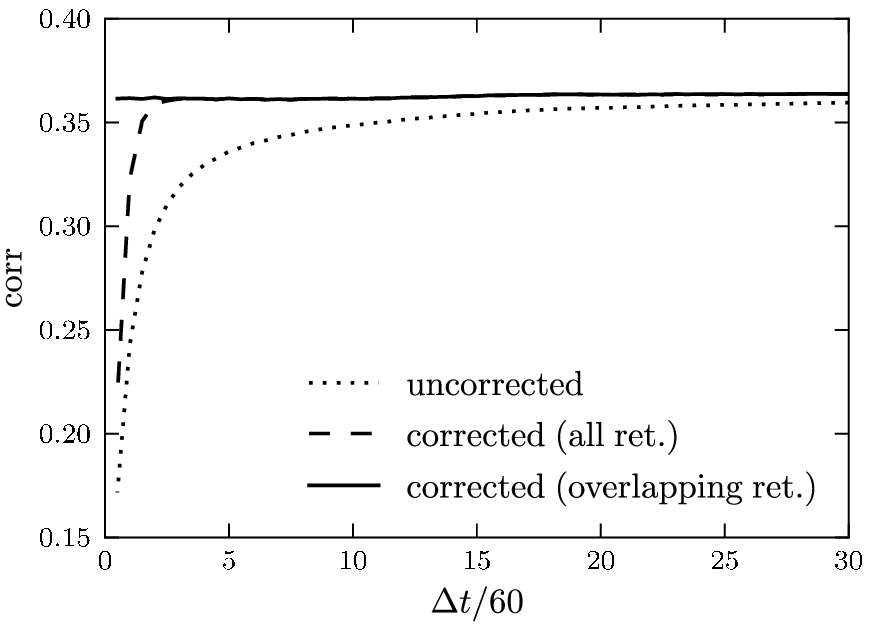}
}
\caption{Compensation of asynchrony effects within the model \uline{with different approaches in the generation of the underlying time series}. The dashed line represents the correlation coefficient, which is corrected by the overlap. The solid line regards in addition only returns, in which time intervals trades occurred.}
\label{fig:ccor}
\end{center}
\end{figure}

$\Delta{t}_{\mathrm{o}}(t)/\Delta{t}$ is the fractional overlap of the corresponding return interval. The fractional overlap does not depend on the actual timescale of the underlying time series.
As equation (\ref{result}) clearly shows, the correlation coefficient of the synchronous part of the return time series is multiplied by the fractional overlap. Hence, this effect can be compensated by
\begin{equation}
\mathrm{corr}_{\mathrm{corrected}}(r_{\Delta t}^{(1)},r_{\Delta t}^{(2)})  = \frac{1}{T}\sum\limits_{j=0}^{T}{g}_{\Delta t}^{(1)}(t_{j}){g}_{\Delta t}^{(2)}(t_{j}) \frac{\Delta{t}}{\Delta{t}_{\mathrm{o}}(t_{j})}\ .
\end{equation}
The dashed line in Fig. \ref{fig:ccor} represents the asynchrony-compensated correlation within our simulation. It turns out that there is a remaining effect that still causes a downscaling of the correlation coefficient for very small return intervals. This behavior occurs when the price of either of the stocks did not change during the return-interval and therefore the corresponding return equals zero. Of course, this event becomes more probable on smaller return intervals $\Delta t$. It corresponds to the small peak at $\Delta t_{\mathrm{o}}=0$ in Fig. \ref{fig:overlapsa}. This remaining downscaling coincides with the cumulative estimator described by Hayashi and Yoshida \cite{hayashi05}. 
It can also be expressed in the formalism used here. It reads
\begin{equation}
\mathrm{corr}(r_{\Delta t}^{(1)},r_{\Delta t}^{(2)})\Big|_{(\gamma^{(1)}_{1}(t)\neq\gamma^{(1)}(t+\Delta t)) \wedge(\gamma^{(2)}(t)\neq\gamma^{(2)}(t+\Delta t))}\ .
\end{equation}
Therefore, when combining both estimators, and thus only regarding returns with overlapping time intervals, the remaining scaling behavior for very small returns can be compensated as well.

As displayed in Fig. \ref{fig:overlaps}, the overlap can also be larger than the actual return interval, implying that terms with such overlaps are corrected downwards. Therefore the compensation can amplify a specific term of the correlation coefficient as well as it can attenuate it.

\section{Application to market data}
\label{application}

Certainly, many aspects contribute to the Epps effect. Our present aim is to quantify the part, which is caused by the asynchrony of the time series.

It is difficult to isolate the Epps Effect on single stock pairs, as it can superimpose with other effects leading to other characteristics of the correlation coefficient  than expected according to the Epps effect. A common approach on this topic is to pick the pairs of stocks, which show a distinctive Epps effect and focus the analysis to these pairs \cite{kwapien04,toth09,voev07}. In the following, we would like to take a different approach:

We classify two ensembles of stock pairs. After compensating the asynchrony effect for each pair, we build the average for the ensemble.
We also plot the error bars representing the double standard deviation $2\sigma$. By this method, we can show the scope of the asynchrony model and identify regions, in which other effects dominate. All data was extracted from the NYSE's TAQ database for the year 2007 \cite{TAQ}. 

The first ensemble consists of stock pairs which provide the most stable correlation. Thereby we want to suppress those effects which are caused by a change in the correlation  during the period in which the correlation coefficient is calculated. \uline{This ensemble represents ideal test conditions for the asynchrony compensation}. To identify those stock pairs with a stable correlation, we calculate the correlation coefficient of 30 daily returns. After shifting this window in 1-day intervals through the year, we calculate the variance of the obtained correlation coefficients ($\mathrm{var_{corr}}$). Then we identify the five stocks providing the smallest variance for each \emph{Global Industry Classification System} (GICS) branch of the \emph{Standard \& Poor's} (S\&P) 500 index. This results in an ensemble of 50 stocks as shown in table \ref{tab:stable}, appendix \ref{ensembles}.

\uline{As the correlation structure of stocks can be non-stationary, we also evaluate the asynchrony compensation without the restriction to stable correlations.} For this purpose, we select a second ensemble consisting of 5 stock pairs of each GICS branch of the S\&P 500 index, whose daily returns are providing the strongest correlation during the year 2007. \uline{These stocks include highly non-stationary correlations as indicated} in table \ref{tab:high}, appendix \ref{ensembles} \uline{(row ``$var_{\mathrm{corr}}$'')}.

\begin{figure}[tb]
\begin{center}
 \subfigure[Ensemble of most stable correlations]{
   \includegraphics[width=0.45\textwidth]{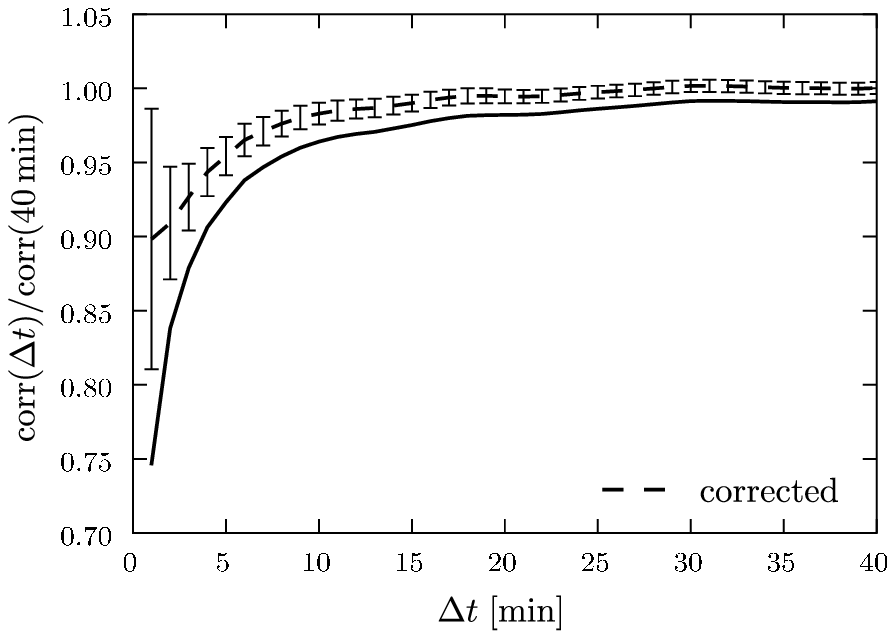}
  }
\subfigure[Ensemble of highest correlations]{
  \includegraphics[width=0.45\textwidth]{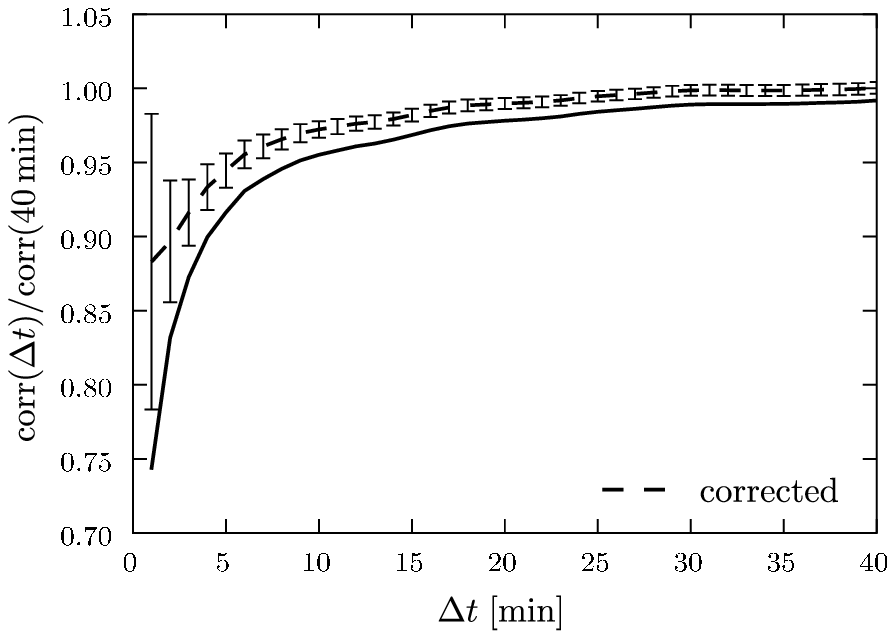}
  }

\caption{Asynchrony-compensated correlations of two ensembles. The data has been normalized to its value at 40 minutes. The error bars represent the double standard deviation.}
\label{fig:ensemble}
\end{center}
\end{figure}

Fig. \ref{fig:ensemble} shows the ensemble average of the correlation coefficient and the asynchrony-compensated correlation coefficient for both ensembles in 2007 (250 trading days).
Before averaging, the correlation coefficients for each stock have been normalized to the value at a return interval $\Delta t = 40$ minutes.

When looking at the whole ensemble we discover that the asynchrony has a pronounced impact on the Epps effect. The asynchrony effect seems to be the dominating cause for the Epps effect on return intervals down to  approximately 10 minutes, where the remaining Epps effect is on average less than 3\% of the correlation coefficient's saturation value at large return intervals. For smaller return intervals, other effects dominate, e.g. a lag between the time series of two stocks, as recent study indicates \cite{toth09}. 

\begin{figure}[tb]
\begin{center}

\subfigure[D - XEL]{
   \includegraphics[width=0.45\textwidth]{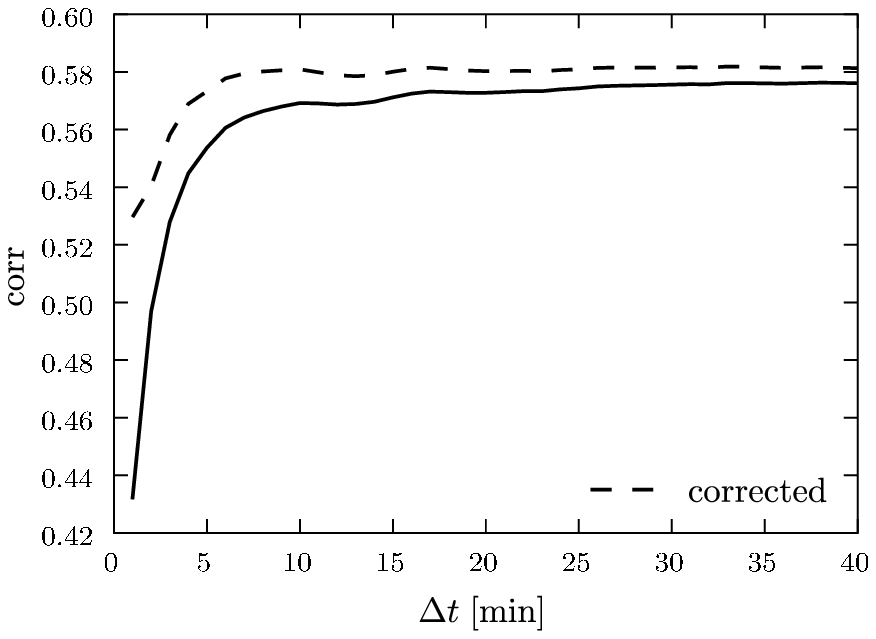}
  \label{fig:realdata_clear}
  }
 \subfigure[AMGN - GENZ]{
  \includegraphics[width=0.45\textwidth]{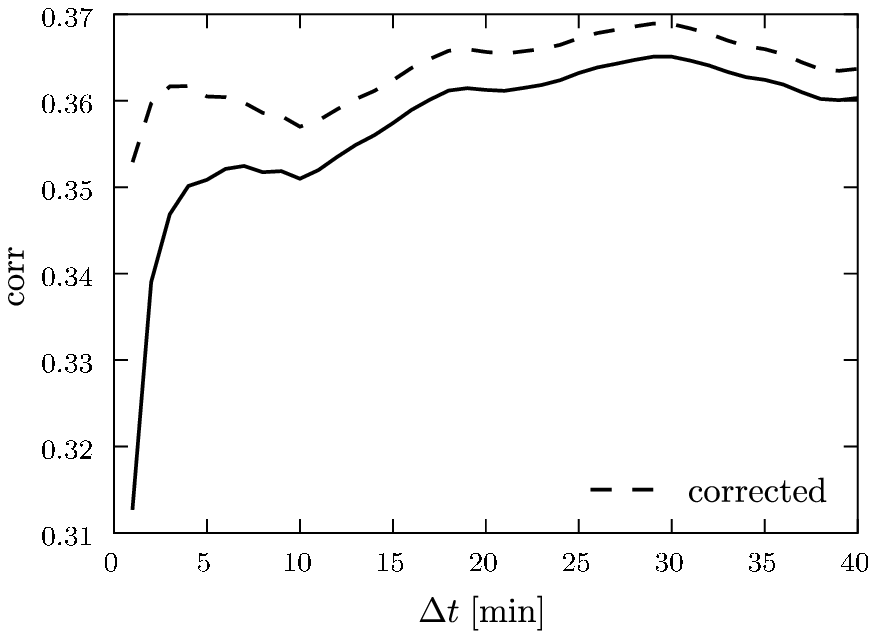}
    \label{fig:realdata_dirty}
  }\\
\caption{Asynchrony-compensated correlations between Dominion Resources, Inc. (D) - Xcel Energy Inc. (XEL) and Amgen Inc. (AMGN) - Genzyme Corp. (GENZ)}
\label{fig:realdata} 
\end{center}
\end{figure}

However, the ensemble consists also of stocks which are very frequently traded, providing a very short average waiting time which results in a fractional overlap $\Delta_{t_{\mathrm{o}}}(t)/\Delta t$ close to unity. Evidently the presented compensation only has a small impact on the correlation estimation of these stocks, as they are so frequently traded that their time series can almost be described as continuous.
Naturally the presented compensation works best for less frequently traded stocks, as they actually show an asynchronous behavior. Fig. \ref{fig:realdata} provides two examples of stock pairs for which the asynchrony of time series is a major effect. While Fig. \ref{fig:realdata_clear} shows a ``clean'' Epps effect, Fig. \ref{fig:realdata_dirty} shows an Epps effect which is superimposed with other phenomena. 

Of course, within the statistical ensemble stock pairs can be found that either do not show an Epps effect or that are so infrequently traded that the assumption of an underlying timeline seems to be unreasonable. Even though the assumption of an underlying time series is a common and intuitive approach on this topic, it may not be valid for very infrequently traded stocks. 

When looking at single stock pairs, it turns out that the asynchrony-compensation works well, if a distinguished Epps effect is found. In case of adopting the presented method as a black box model without looking at the scaling behavior of the correlation coefficient, we believe that a return interval of 5 minutes represents a good lower bound for the scope of this method.

\section{Conclusion}
\label{conclusion}
We presented a model for the scaling behavior of financial correlations due to the asynchrony of the time series. This purely statistical effect can be compensated. Furthermore, we applied this compensation to market data under the assumption of an underlying time series with non-lagged correlations. We quantified the influence of the asynchrony on the overall decay of the correlation coefficient towards small return intervals, which is known as the Epps effect. The results clearly demonstrate that the asynchrony can have a huge impact on the Epps effect. It rather can be the dominating cause for less frequently traded stocks. The main advantage of our method is that no parameters or adjustments are necessary, since it is based on the trading times only.

In our empirical study, the asynchrony-compensation allowed us to recover the correlation coefficient for return intervals down to 10 minutes. At this return interval, the remaining Epps effect is on average less than 3\% of the correlation coefficient's saturation value at large return intervals. \uline{We also demonstrated that the presented method holds for non-stationary correlated time series.} The accurate calculation of correlations is of major importance for risk management. To keep the estimation error small, a long time series of returns is required. Yet at the same time, the time series should not reach too far into the past. The latter is important because the correlation structure can be highly dynamic, as the dramatic events of autumn 2008 prove.
Applying our method to intraday data allows to choose smaller return intervals and hence provides improved statistical significance of the correlations for the same time horizon.

\section*{Acknowledgements}
The authors thank J. H\"ammerling and V. Osipov for fruitful discussion.
M.C.M. acknowledges financial support from Studienstiftung des deutschen Volkes.

\begin{appendix}
\section{Relation between $g^{(i)}_{\Delta t}$ and $\tilde{g}^{(i)}$}
\label{app1}
We defined the normalzed returns as
\begin{eqnarray}
g^{(i)}_{\Delta t}(t)=\frac{r^{(i)}_{\Delta t}(t)-\langle r^{(i)}_{\Delta t}  \rangle}{\sqrt{\mathrm{Var}(r^{(i)}_{\Delta t})}} \label{anormret}
\\
\tilde{g}^{(i)}(t)=\frac{\tilde{r}^{(i)}(t)-\langle \tilde{r}^{(i)}  \rangle}{\sqrt{\mathrm{Var}(\tilde{r}^{(i)})}} \label{anormretu}\ ,
\end{eqnarray}
where $\mathrm{Var}(\cdots)$ refers to the variance of a time series.
Inserting the return, expressed through the underlying time series,
\begin{equation}
r^{(i)}(t) = \sum\limits_{j=0}^{N^{(i)}_{\Delta t}(t)}{\tilde{r}^{(i)}(\gamma^{(i)}(t)+j\Delta \tilde{t}})\ ,
\end{equation}
in equation (\ref{anormret}), results in

\begin{eqnarray}
g^{(i)}_{\Delta t}(t)&=&\frac{\sum\limits_{j=0}^{N^{(i)}_{\Delta t}(t)}{\left(\tilde{r}^{(i)}(\gamma^{(i)}(t)+j\Delta \tilde{t})\right)}-\langle r^{(i)}_{\Delta t}  \rangle}{\sqrt{\mathrm{Var}(r^{(i)}_{\Delta t})}}\\
&=&\frac{\sqrt{\mathrm{Var}(\tilde{r}^{(i)})}\sum\limits_{j=0}^{N^{(i)}_{\Delta t}(t)}{\left(\tilde{g}^{(i)}(\gamma^{(i)}(t)+j\Delta \tilde{t})\right)}-\langle r^{(i)}_{\Delta t}  \rangle+N^{(i)}_{\Delta t}(t)\langle \tilde{r}^{(i)}  \rangle}{\sqrt{\mathrm{Var}(r^{(i)}_{\Delta t})}}\ . \label{ainterm}
\end{eqnarray}
In equation (\ref{ainterm}), (\ref{anormretu}) was used to express the underlying returns $\tilde{r}^{(i)}$.\\
The mean values and variance are additive, which leads to 

\begin{equation}
\left\langle r^{(i)}_{\Delta t} \right\rangle = \left\langle N^{(i)}_{\Delta t}(t) \right\rangle \left\langle \tilde{r}^{(i)}  \right\rangle
,\quad
\mathrm{Var}(r^{(i)}_{\Delta t}) = \left\langle N^{(i)}_{\Delta t}(t) \right\rangle \mathrm{Var}( \tilde{r}^{(i)})\ .
\end{equation}
Therefore, we obtain

\begin{equation}
g_{\Delta t}^{(i)}(t) = \frac{1}{\sqrt{\left\langle N^{(i)}_{\Delta t}(t)\right\rangle}} \sum\limits_{j=0}^{N^{(i)}_{\Delta t}(t))}{\tilde{g}^{(i)}(\gamma^{(i)}(t)+j\Delta \tilde{t}})
- \frac{\langle \tilde{r}^{(i)} \rangle \left(\left\langle N^{(i)}_{\Delta t}(t)\right\rangle- N^{(i)}_{\Delta t}(t)\right)}{\sqrt{\mathrm{Var}(r^{(i)}_{\Delta t})}} \ .
\end{equation}
As the average time interval per return converges to $\Delta t$, the mean number of underlying price changes $\left\langle N^{(i)}_{\Delta t}(t)\right\rangle$ is given by $\Delta t / \Delta \tilde{t}$. Thus, we arrive at

\begin{equation}
g_{\Delta t}^{(i)}(t) = \sqrt{\frac{\Delta \tilde{t}}{\Delta t}}  \sum\limits_{j=0}^{N^{(i)}_{\Delta t}(t))}{\tilde{g}^{(i)}(\gamma^{(i)}(t)+j\Delta \tilde{t}})
- \frac{\langle \tilde{r}^{(i)} \rangle \left( \frac{\Delta t}{\Delta \tilde{t}} - N^{(i)}_{\Delta t}(t)\right)}{\sqrt{\mathrm{Var}(r^{(i)}_{\Delta t})}} \ ,
\end{equation}
which is equation (\ref{eq:appneeded}).
\clearpage

\begin{table}[b]
\section{Stock ensembles}
\label{ensembles}

\caption{Top 5 five stock pairs with the most stable correlation from each GICS branch of the S\&P 500 index.}
\label{tab:stable}
{\tiny

\begin{tabular}{| l | l | l | l | l | l | l | l | l | c | c |}
\hline
GICS Branch & \multicolumn{4}{|c|}{Stock 1} & \multicolumn{4}{|c|}{Stock 2} & $\mathrm{corr}$ & $\mathrm{var_{corr}}$ \\
 & Symbol & Name & Stock Exchange & Volume & Symbol & Name & Stock Exchange & Volume & & \\
\hline \multirow{5}{*}{Consumer Discretionary}
& AMZN & Amazon Corp. & NASDAQ & 1215700 & SBUX & Starbucks Corp. & NASDAQ & 1287200 & 0.80 & 0.45\\
& APOL & Apollo Group & NASDAQ & 332200 & SHLD & Sears Holdings Corporation & NASDAQ & 317900 & 0.28 & 0.57\\
& AMZN & Amazon Corp. & NASDAQ & 1215700 & SPLS & Staples Inc. & NASDAQ & 674500 & 0.80 & 0.62\\
& AMZN & Amazon Corp. & NASDAQ & 1215700 & CMCSA & Comcast Corp. & NASDAQ & 910000 & 0.69 & 0.64\\
& EXPE & Expedia Inc. & NASDAQ & 883300 & SHLD & Sears Holdings Corporation & NASDAQ & 317900 & 0.43 & 0.66\\
\hline \multirow{5}{*}{Consumer Staples}
& COST & Costco Co. & NASDAQ & 237500 & PG & Procter \& Gamble & NYSE & 1396077700 & 0.05 & 0.71\\
& COST & Costco Co. & NASDAQ & 237500 & CVS & CVS Caremark Corp. & NYSE & 1454768200 & 0.09 & 0.77\\
& KO & Coca Cola Co. & NYSE & 1152559200 & COST & Costco Co. & NASDAQ & 237500 & 0.04 & 0.79\\
& MO & Altria Group  Inc. & NYSE & 1312801100 & CCE & Coca-Cola Enterprises & NYSE & 335449400 & 0.29 & 0.79\\
& CCE & Coca-Cola Enterprises & NYSE & 335449400 & KFT & Kraft Foods Inc-A & NYSE & 1438368700 & 0.32 & 0.80\\
\hline \multirow{5}{*}{Energy}
& EP & El Paso Corp. & NYSE & 697103700 & SE & Spectra Energy Corp. & NYSE & 331523100 & 0.34 & 0.84\\
& CVX & Chevron Corp. & NYSE & 1271849400 & SE & Spectra Energy Corp. & NYSE & 331523100 & 0.26 & 0.85\\
& HES & Hess Corporation & NYSE & 427802700 & SE & Spectra Energy Corp. & NYSE & 331523100 & 0.23 & 0.85\\
& MUR & Murphy Oil & NYSE & 223916300 & SE & Spectra Energy Corp. & NYSE & 331523100 & 0.25 & 0.86\\
& SII & Smith International & NYSE & 370893400 & SE & Spectra Energy Corp. & NYSE & 331523100 & 0.32 & 0.86\\
\hline \multirow{5}{*}{Financials}
& SCHW & Charles Schwab & NASDAQ & 1445500 & ETFC & E*Trade Financial Corp. & NASDAQ & 1391800 & 0.67 & 0.44\\
& ETFC & E*Trade Financial Corp. & NASDAQ & 1391800 & FITB & Fifth Third Bancorp & NASDAQ & 218000 & 0.29 & 0.58\\
& SCHW & Charles Schwab & NASDAQ & 1445500 & FITB & Fifth Third Bancorp & NASDAQ & 218000 & 0.36 & 0.58\\
& SCHW & Charles Schwab & NASDAQ & 1445500 & HCBK & Hudson City Bancorp & NASDAQ & 686900 & 0.54 & 0.60\\
& ACAS & American Capital Strategies Ltd & NASDAQ & 207200 & SCHW & Charles Schwab & NASDAQ & 1445500 & 0.50 & 0.60\\
\hline \multirow{5}{*}{Health Care}
& CELG & Celgene Corp. & NASDAQ & 619200 & ESRX & Express Scripts & NASDAQ & 998400 & 0.65 & 0.47\\
& AMGN & Amgen & NASDAQ & 813900 & CELG & Celgene Corp. & NASDAQ & 619200 & 0.48 & 0.50\\
& AMGN & Amgen & NASDAQ & 813900 & BIIB & BIOGEN IDEC Inc. & NASDAQ & 381800 & 0.48 & 0.52\\
& CELG & Celgene Corp. & NASDAQ & 619200 & THC & Tenet Healthcare Corp. & NYSE & 805228900 & -0.23 & 0.53\\
& AMGN & Amgen & NASDAQ & 813900 & GENZ & Genzyme Corp. & NASDAQ & 242900 & 0.57 & 0.53\\
\hline \multirow{5}{*}{Industrials}
& GE & General Electric & NYSE & 4303823300 & LUV & Southwest Airlines & NYSE & 862775700 & 0.53 & 0.73\\
& MMM & 3M Company & NYSE & 549124400 & CBE & Cooper Industries  Ltd. & NYSE & 175911300 & 0.24 & 0.73\\
& CBE & Cooper Industries  Ltd. & NYSE & 175911300 & GWW & Grainger (W.W.) Inc. & NYSE & 95324400 & 0.15 & 0.73\\
& CBE & Cooper Industries  Ltd. & NYSE & 175911300 & GR & Goodrich Corporation & NYSE & 144177800 & 0.15 & 0.75\\
& CBE & Cooper Industries  Ltd. & NYSE & 175911300 & FLR & Fluor Corp. (New) & NYSE & 171713100 & 0.10 & 0.76\\
\hline \multirow{5}{*}{Information Technology}
& AAPL & Apple Inc. & NASDAQ & 4627500 & INTC & Intel Corp. & NASDAQ & 5529100 & 0.84 & 0.19\\
& AAPL & Apple Inc. & NASDAQ & 4627500 & CSCO & Cisco Systems & NASDAQ & 4886800 & 0.76 & 0.24\\
& AAPL & Apple Inc. & NASDAQ & 4627500 & YHOO & Yahoo Inc. & NASDAQ & 2609700 & 0.71 & 0.25\\
& AAPL & Apple Inc. & NASDAQ & 4627500 & ORCL & Oracle Corp. & NASDAQ & 1731900 & 0.78 & 0.25\\
& AAPL & Apple Inc. & NASDAQ & 4627500 & EBAY & eBay Inc. & NASDAQ & 1056100 & 0.73 & 0.26\\
\hline \multirow{5}{*}{Materials}
& MON & Monsanto Co. & NYSE & 479605700 & SEE & Sealed Air Corp.(New) & NYSE & 126716200 & 0.06 & 0.52\\
& FCX & Freeport-McMoran Cp \& Gld & NYSE & 1058215000 & SIAL & Sigma-Aldrich & NASDAQ & 133800 & 0.05 & 0.54\\
& ECL & Ecolab Inc. & NYSE & 163404500 & SEE & Sealed Air Corp.(New) & NYSE & 126716200 & 0.26 & 0.60\\
& ATI & Allegheny Technologies Inc & NYSE & 269746100 & SEE & Sealed Air Corp.(New) & NYSE & 126716200 & 0.10 & 0.63\\
& PX & Praxair  Inc. & NYSE & 245761900 & SEE & Sealed Air Corp.(New) & NYSE & 126716200 & 0.12 & 0.65\\
\hline \multirow{5}{*}{Telecommunication Services}
& Q & Qwest Communications Int & NYSE & 1623807700 & S & Sprint Nextel Corp. & NYSE & 2044634000 & 0.52 & 0.84\\
& Q & Qwest Communications Int & NYSE & 1623807700 & VZ & Verizon Communications & NYSE & 1472335800 & 0.49 & 0.86\\
& S & Sprint Nextel Corp. & NYSE & 2044634000 & VZ & Verizon Communications & NYSE & 1472335800 & 0.49 & 0.87\\
& AMT & American Tower Corp. & NYSE & 387199400 & Q & Qwest Communications Int & NYSE & 1623807700 & 0.26 & 0.97\\
& AMT & American Tower Corp. & NYSE & 387199400 & WIN & Windstream Corporation & NYSE & 400634200 & 0.10 & 0.99\\
\hline \multirow{5}{*}{Utilities}
& DUK & Duke Energy & NYSE & 902519200 & DYN & Dynegy Inc. & NYSE & 702035600 & 0.54 & 0.74\\
& CMS & CMS Energy & NYSE & 264225200 & DYN & Dynegy Inc. & NYSE & 702035600 & 0.48 & 0.79\\
& CMS & CMS Energy & NYSE & 264225200 & DUK & Duke Energy & NYSE & 902519200 & 0.39 & 0.81\\
& AES & AES Corp. & NYSE & 556049300 & CMS & CMS Energy & NYSE & 264225200 & 0.31 & 0.84\\
& CNP & CenterPoint Energy & NYSE & 359757800 & DUK & Duke Energy & NYSE & 902519200 & 0.33 & 0.84\\
\hline\end{tabular}

}
\end{table}

\begin{table}[h]
\caption{Top 5 five stock pairs with the highest correlation from each GICS branch of the S\&P 500 index.}
\label{tab:high}
{\tiny

\begin{tabular}{| l | l | l | l | l | l | l | l | l | c | c |}
\hline
GICS Branch & \multicolumn{4}{|c|}{Stock 1} & \multicolumn{4}{|c|}{Stock 2} & $\mathrm{corr}$ & $\mathrm{var_{corr}}$ \\
 & Symbol & Name & Stock Exchange & Volume & Symbol & Name & Stock Exchange & Volume & & \\
\hline \multirow{5}{*}{Consumer Discretionary}
& APOL & Apollo Group & NASDAQ & 332200 & SPLS & Staples Inc. & NASDAQ & 674500 & 0.77 & 1.08\\
& BBBY & Bed Bath \& Beyond & NASDAQ & 233000 & SPLS & Staples Inc. & NASDAQ & 674500 & 0.79 & 1.08\\
& AMZN & Amazon Corp. & NASDAQ & 1215700 & SPLS & Staples Inc. & NASDAQ & 674500 & 0.80 & 0.62\\
& AMZN & Amazon Corp. & NASDAQ & 1215700 & SBUX & Starbucks Corp. & NASDAQ & 1287200 & 0.80 & 0.45\\
& SPLS & Staples Inc. & NASDAQ & 674500 & SBUX & Starbucks Corp. & NASDAQ & 1287200 & 0.81 & 0.69\\
\hline \multirow{5}{*}{Consumer Staples}
& KO & Coca Cola Co. & NYSE & 1152559200 & SLE & Sara Lee Corp. & NYSE & 542934700 & 0.42 & 0.90\\
& SLE & Sara Lee Corp. & NYSE & 542934700 & WMT & Wal-Mart Stores & NYSE & 1992433400 & 0.43 & 1.02\\
& CVS & CVS Caremark Corp. & NYSE & 1454768200 & KFT & Kraft Foods Inc-A & NYSE & 1438368700 & 0.44 & 0.98\\
& KFT & Kraft Foods Inc-A & NYSE & 1438368700 & SLE & Sara Lee Corp. & NYSE & 542934700 & 0.48 & 0.87\\
& COST & Costco Co. & NASDAQ & 237500 & WFMI & Whole Foods Market & NASDAQ & 211400 & 0.59 & 1.09\\
\hline \multirow{5}{*}{Energy}
& EP & El Paso Corp. & NYSE & 697103700 & RDC & Rowan Cos. & NYSE & 369346200 & 0.39 & 0.91\\
& CVX & Chevron Corp. & NYSE & 1271849400 & XOM & Exxon Mobil Corp. & NYSE & 2798325600 & 0.41 & 1.00\\
& XOM & Exxon Mobil Corp. & NYSE & 2798325600 & HAL & Halliburton Co. & NYSE & 1701703200 & 0.45 & 1.18\\
& CVX & Chevron Corp. & NYSE & 1271849400 & EP & El Paso Corp. & NYSE & 697103700 & 0.45 & 0.88\\
& COP & ConocoPhillips & NYSE & 1382115600 & EP & El Paso Corp. & NYSE & 697103700 & 0.46 & 0.94\\
\hline \multirow{5}{*}{Financials}
& CINF & Cincinnati Financial & NASDAQ & 55900 & TROW & T. Rowe Price Group & NASDAQ & 216400 & 0.60 & 1.74\\
& FITB & Fifth Third Bancorp & NASDAQ & 218000 & HBAN & Huntington Bancshares & NASDAQ & 71600 & 0.62 & 2.01\\
& FITB & Fifth Third Bancorp & NASDAQ & 218000 & TROW & T. Rowe Price Group & NASDAQ & 216400 & 0.63 & 1.34\\
& SCHW & Charles Schwab & NASDAQ & 1445500 & ETFC & E*Trade Financial Corp. & NASDAQ & 1391800 & 0.67 & 0.44\\
& TROW & T. Rowe Price Group & NASDAQ & 216400 & ZION & Zions Bancorp & NASDAQ & 48500 & 0.70 & 1.40\\
\hline \multirow{5}{*}{Health Care}
& BIIB & BIOGEN IDEC Inc. & NASDAQ & 381800 & PDCO & Patterson Cos. Inc. & NASDAQ & 106400 & 0.60 & 1.15\\
& BIIB & BIOGEN IDEC Inc. & NASDAQ & 381800 & GENZ & Genzyme Corp. & NASDAQ & 242900 & 0.62 & 0.85\\
& GENZ & Genzyme Corp. & NASDAQ & 242900 & GILD & Gilead Sciences & NASDAQ & 275100 & 0.64 & 0.91\\
& CELG & Celgene Corp. & NASDAQ & 619200 & ESRX & Express Scripts & NASDAQ & 998400 & 0.65 & 0.47\\
& BSX & Boston Scientific & NYSE & 1205569800 & THC & Tenet Healthcare Corp. & NYSE & 805228900 & 0.68 & 0.60\\
\hline \multirow{5}{*}{Industrials}
& GE & General Electric & NYSE & 4303823300 & LUV & Southwest Airlines & NYSE & 862775700 & 0.53 & 0.73\\
& CTAS & Cintas Corporation & NASDAQ & 48500 & EXPD & Expeditors Int'l & NASDAQ & 215600 & 0.54 & 1.48\\
& EXPD & Expeditors Int'l & NASDAQ & 215600 & MNST & Monster Worldwide & NASDAQ & 196700 & 0.55 & 1.20\\
& CHRW & C.H. Robinson Worldwide & NASDAQ & 112400 & EXPD & Expeditors Int'l & NASDAQ & 215600 & 0.56 & 1.46\\
& MNST & Monster Worldwide & NASDAQ & 196700 & PCAR & PACCAR Inc. & NASDAQ & 164400 & 0.56 & 1.23\\
\hline \multirow{5}{*}{Information Technology}
& DELL & Dell Inc. & NASDAQ & 909400 & ORCL & Oracle Corp. & NASDAQ & 1731900 & 0.79 & 0.47\\
& CSCO & Cisco Systems & NASDAQ & 4886800 & DELL & Dell Inc. & NASDAQ & 909400 & 0.79 & 0.41\\
& INTC & Intel Corp. & NASDAQ & 5529100 & ORCL & Oracle Corp. & NASDAQ & 1731900 & 0.82 & 0.29\\
& CSCO & Cisco Systems & NASDAQ & 4886800 & ORCL & Oracle Corp. & NASDAQ & 1731900 & 0.83 & 0.34\\
& AAPL & Apple Inc. & NASDAQ & 4627500 & INTC & Intel Corp. & NASDAQ & 5529100 & 0.84 & 0.19\\
\hline \multirow{5}{*}{Materials}
& DD & Du Pont (E.I.) & NYSE & 716944300 & FCX & Freeport-McMoran Cp \& Gld & NYSE & 1058215000 & 0.36 & 1.02\\
& FCX & Freeport-McMoran Cp \& Gld & NYSE & 1058215000 & MON & Monsanto Co. & NYSE & 479605700 & 0.36 & 0.85\\
& APD & Air Products \& Chemicals & NYSE & 187953500 & BLL & Ball Corp. & NYSE & 119560600 & 0.36 & 0.88\\
& ECL & Ecolab Inc. & NYSE & 163404500 & NEM & Newmont Mining Corp. (Hldg. Co.) & NYSE & 958900000 & 0.37 & 0.93\\
& FCX & Freeport-McMoran Cp \& Gld & NYSE & 1058215000 & NEM & Newmont Mining Corp. (Hldg. Co.) & NYSE & 958900000 & 0.43 & 1.06\\
\hline \multirow{5}{*}{Telecommunication Services}
& T & AT\&T Inc. & NYSE & 2663617200 & Q & Qwest Communications Int & NYSE & 1623807700 & 0.45 & 1.01\\
& S & Sprint Nextel Corp. & NYSE & 2044634000 & VZ & Verizon Communications & NYSE & 1472335800 & 0.49 & 0.87\\
& Q & Qwest Communications Int & NYSE & 1623807700 & VZ & Verizon Communications & NYSE & 1472335800 & 0.49 & 0.86\\
& T & AT\&T Inc. & NYSE & 2663617200 & VZ & Verizon Communications & NYSE & 1472335800 & 0.50 & 1.05\\
& Q & Qwest Communications Int & NYSE & 1623807700 & S & Sprint Nextel Corp. & NYSE & 2044634000 & 0.52 & 0.84\\
\hline \multirow{5}{*}{Utilities}
& DUK & Duke Energy & NYSE & 902519200 & TE & TECO Energy & NYSE & 177983100 & 0.35 & 0.87\\
& D & Dominion Resources & NYSE & 321656100 & XEL & Xcel Energy Inc & NYSE & 337262500 & 0.36 & 0.92\\
& CMS & CMS Energy & NYSE & 264225200 & DUK & Duke Energy & NYSE & 902519200 & 0.39 & 0.81\\
& CMS & CMS Energy & NYSE & 264225200 & DYN & Dynegy Inc. & NYSE & 702035600 & 0.48 & 0.79\\
& DUK & Duke Energy & NYSE & 902519200 & DYN & Dynegy Inc. & NYSE & 702035600 & 0.54 & 0.74\\
\hline\end{tabular}

}
\end{table}
\end{appendix}
\clearpage
\bibliographystyle{elsarticle-num}
\bibliography{Manuskript_v0.7.bbl}
\end{document}